\documentclass[aps,showpacs,preprintnumbers,amsmath, amssymb]{revtex4}

\usepackage{float}
\usepackage{graphics,epsfig}
\usepackage{graphicx}
\usepackage{dcolumn}
\usepackage{bm}
\usepackage{slashbox}
\usepackage{CJK}
\begin{document}
\begin{CJK}{GBK}{song}
\baselineskip=0.8 cm
\title{{\bf Gauss-Bonnet holographic superconductors in Born-Infeld electrodynamics with backreactions}}
\author{Yunqi Liu, Yan Peng, Bin Wang}
\affiliation{INPAC and Department of Physics, Shanghai Jiao Tong
University, Shanghai 200240, China}

\vspace*{0.2cm}
\begin{abstract}
\baselineskip=0.6 cm
\begin{center}
{\bf Abstract}
\end{center}

We develop a general holographic superconductor
away from the probe limit by considering the
corrections both in the gravity and in the gauge matter
fields.  We find the consistent effects of the
high curvature correction in the gravity, the
nonlinear correction in the gauge matter field
and the backreaction on the dynamics of bulk AdS
background and boundary CFT. They all have the effect to protect
the stability of the bulk spacetime and hinder
the formation of the scalar hair condensation on
the boundary.

\end{abstract}

\maketitle
\newpage
\vspace*{0.2cm}

\section{Introduction}
The holographic model of superconductors, which
is constructed by a gravitational theory of a
Maxwell field coupled to a charged complex scalar
field via anti-de Sitter/conformal field theory
(AdS/CFT) correspondence, has been investigated
extensively in the past years (for reviews, see
Refs. \cite{HorowitzRev}-\cite{HerzogRev} and
references therein). According to the AdS/CFT
correspondence, the emergence of the scalar hair
in the bulk AdS black hole corresponds to the
formation of a charged condensation in the
boundary dual CFTs. This provides a novel way to
investigate the superconductor and attracts
considerable interest for its potential
applications to the condensed matter physics
\cite{gubser,3h2,HorowitzPRD78,Nakano-Wen,Amado,Koutsoumbas,
Umeh,Sonner,Jing-Chen,Franco,Herzog-2010,maeda,gregorysoda,Konoplya,Siopsis,CaiNie,Ge-Wang-Wu}.

Motivated by the application of the Mermin-Wagner theorem to the
holographic superconductors, there have been a lot of interest in
generalizing the Einstein gravity background in the holographic
superconductor to include the curvature correction in gravity. By examining the
charged scalar field together with a Maxwell field in the
Gauss-Bonnet-AdS black hole background
\cite{gregorysoda,maeda,Pan-Wang,panwang,Barclay-Gregory,siani,jingpanl,Gregory,barcly,cainiezhang,licaizhang,kanno,panjingwang,Setare},
it has been observed that the higher curvature correction makes the
condensation harder to form. Other attempts to study the holographic
superconductor by modifying gravity background can be found in
\cite{cai1,momeni,wucao}.

In the low-energy limit of heterotic string theory, the higher-order
correction term also appears in the Maxwell gauge field
\cite{Gross}. Besides the curvature correction to the gravity, it is
also interesting to investigate the high-order correction related to
the gauge matter field. In the study of the ratio of shear viscosity
to entropy density, it was found that the higher derivative
correction to the gravity has different effect from that of the
higher derivative correction to the gauge matter fields
\cite{caisun}. This motivates people to study the effect of higher
order corrected Maxwell field on the scalar condensation and compare
with the effect brought by the correction in the gravity
\cite{panjingwang1}. Considering that the Born-Infeld
electrodynamics \cite{born} is the only possible non-linear version
of electrodynamics which is invariant under electromagnetic duality
transformations \cite{gibbons}, the nonlinear extension on the gauge
field described by the Born-Infeld electrodynamics has been employed
in the study on the stability of the bulk spacetime \cite{liuwang}
and its influence on the condensation in CFT
\cite{CaiNie,jingpanchen,jingpanchen1,Gangopadhyay}. These studies
were concentrated on the background of neutral AdS black holes and
in the probe limit approximation.

In the present work we would like to combine the higher order
gravity corrections together with the higher corrections to the
gauge matter fields in the study of the holographic superconductor
and compare their effects on the condensation. We will examine the
holographic superconductor away from the probe approximation and
take the backreaction of the spacetime into account. Considering the
backreaction of classical fields onto the spacetime is not trivial.
In \cite{liuwang} it was found that the nonzero backreaction can
make the neutral AdS black hole become Born-Infeld AdS black hole.
In our work, considering the backreaction, we will obtain the
Gauss-Bonnet Born-Infeld AdS black hole in the gravity side and we
will examine the condensation of the scalar field in this background
when it is coupled with the Born-Infeld electromagnetic field. We
will present a complete picture of the holographic superconductor
with corrections in the gravity side together with the gauge matter
fields and take the backreaction into account.

The outline of this work is as follows. In section II, we will
introduce the model, derive the set of equations of motion. In
section III we will study the dynamics of a scalar perturbation in
the high temperature phase. In section IV we will show the effect of
the backreaction, Gauss-Bonnet factor and the Born-Infeld factor on
the condensate. In section V, we will analyze the conductivity of
the holographic superconductor. We will conclude our main results in
the last section.

\section{Action and the equations of motion}

The general action describing the Born-Infeld
field and a charged complex scalar field in the
five-dimensional Einstein-Gauss-Bonnet spacetime
with negative cosmological constant reads
\begin{eqnarray}\label{action}
S&=&\frac{1}{2\kappa^2}\int d^5
x\sqrt{-g}\left[R+\frac{12}{l^2}+\frac{\alpha}{2}
(R^{abcd}R_{abcd}-4R^{ab}R_{ab}+R^2)\right]\nonumber\\
&+&\int d^5 x\sqrt{-g}\left[\frac{1}{b^2}\left(1-\sqrt{1+\frac{b^2
F^{ab}F_{ab}}{2}}\right)+(-|\nabla\Psi-i q A\Psi
|^2-m^2|\Psi|^2)\right],
\end{eqnarray}
where $\kappa$ is the five dimensional
gravitational constant $\kappa^2=8 \pi G_5$ with
$G_5$ the five-dimensional Newton constant, ~$g$
is the determinant of the metric, ~$l$ here is
the AdS radius, $q$ and $m$ are respectively the
charge and the mass of the scalar field. $b$ is
the Born-Infeld coupling parameter. In the limit
$b\rightarrow0$, the Born-Infeld field will
reduce to the Maxwell field.

When there is no backreaction of classical fields onto the
spacetime, we have the Gauss-Bonnet AdS black hole on the gravity
background with the metric
\begin{eqnarray}
ds^2=-f(r)dt^2+\frac{dr^2}{f(r)}+\frac{r^2}{l_e^2}(dx^2+dy^2+dz^2)~,
\end{eqnarray}
where
\begin{eqnarray}
f(r)=\frac{r^2}{2\alpha}\left[1-\sqrt{1-\frac{4\alpha}{l^2}(1-\frac{r_h^4}{r^4})},
\right]
\end{eqnarray}
and
\begin{eqnarray}
l_e^2=\frac{l^2}{2}\left[1+\sqrt{1-\frac{4\alpha}{l^2}}
\right]
\end{eqnarray}
is the effective AdS radius. The Hawking
temperature of the Gauss-Bonnet AdS black hole,
which will be interpreted as the temperature of
the CFT, can be expressed as
\begin{eqnarray}\label{temperature}
T=\left.\frac{f'(r)}{4 \pi}\right|_{r=r_h}~.
\end{eqnarray}

We will consider the backreaction in our work,
thus we take a metric ansatz of the gravitational
background in the form
\begin{eqnarray}\label{ansatz}
ds^2=-f(r)e^{-\chi(r)}dt^2+\frac{dr^2}{f(r)}+\frac{r^2}{l_e^2}(dx^2+dy^2+dz^2)~.
\end{eqnarray}
The electromagnetic field and the scalar field
can be chosen as
\begin{eqnarray}
A_t=\phi(r)dt,~~~~\Psi=\Psi(r),
\end{eqnarray}
where $\Psi(r)$ can be taken to be real without
loss of generality. The Hawking temperature of
this black hole is changed to be
\begin{eqnarray}\label{temperature}
T=\left.\frac{f'(r)e^{-\chi(r)/2}}{4
\pi}\right|_{r=r_h}~.
\end{eqnarray}

Considering the ansatz of the metric when we take into account the
backreaction, the equations of motions can be obtained
\begin{eqnarray}\label{equationsofmotion}
0&=&\psi ''(r)+\psi '(r)
\left[\frac{3}{r}+\frac{f'(r)}{f(r)}-\frac{\chi '(r)}{2}\right]+\psi
(r) \left[\frac{q^2 \phi (r)^2 e^{\chi
(r)}}{f(r)^2}-\frac{m^2}{f(r)}\right],\nonumber\\
0&=&\left[\phi ''(r)+\left(\frac{\chi'(r)}{2}+\frac{3}{r}\right)\phi
'(r)\right](1-b^2 e^{\chi (r)}\phi '(r)^2 )+\frac{b^2}{2}\phi
'(r)e^{\chi
(r)}\left[\chi'(r)\phi '(r)^2\right.\nonumber\\&&\left.+2 \phi '(r) \phi''(r)\right)]-\frac{2 q^2 \phi (r)\psi (r)^2}{f(r)}(1-b^2 e^{\chi (r)}\phi '(r)^2 )^{\frac{3}{2}}~,\nonumber\\
0&=&(1-\frac{2 \alpha f(r)}{r^2})\chi '(r)+ \frac{4}{3} \kappa ^2 r
\left[\frac{q^2 \phi (r)^2 \psi (r)^2 e^{\chi
(r)}}{f(r)^2}+\psi '(r)^2\right],\nonumber\\
0&=&(1-\frac{2 \alpha
f(r)}{r^2})f'(r)+\frac{2}{r}f(r)-\frac{4r}{l^2}+\frac{2}{3} \kappa
^2 r \left[\left(\frac{q^2 \phi (r)^2 \psi (r)^2 e^{\chi
(r)}}{f(r)}+f(r) \psi '(r)^2  +m^2 \psi
(r)^2\right.\right.\nonumber\\&&\left.\left.\frac{e^{\chi (r)} q^2
\phi(r)^2 \psi(r)^2}{f(r)}\right)+\frac{1}{b^2}\left[\left(1-b^2
\phi'(r)^2\right)^{-1/2}-1\right]\right],
\end{eqnarray}
where the prime denotes the derivative with respect to $r$.

There are two branches of solutions to the above
equations Eq.(\ref{equationsofmotion}), which are
classified into high temperature solutions and
low temperature solutions. In the following we
will provide detailed discussions on these two
branches of solutions.

\section{Dynamics at high temperature phase}

In this part, we will mainly discuss the
solutions of the above equations of motions when
the black hole temperature is high.

Using the separation $\psi_{\omega,~k}(r)e^{-i(\omega t+kx)}$, we
can rewrite the perturbation equation of the scalar field with mass
in the background (\ref{ansatz}) into
\begin{eqnarray}\label{phifield}
0&=&\psi_{\omega,k} ''(r)+\psi_{\omega,k} '(r)
\left[\frac{3}{r}+\frac{f'(r)}{f(r)}-\frac{\chi
'(r)}{2}\right]+\psi_{\omega,k}(r)\left\{\frac{[\omega
+q \phi(r)]^2
e^{\chi(r)}}{f(r)^2}-\frac{m^2}{f(r)}-\frac{l^2
k^2}{f(r) r^2}\right\}~.
\end{eqnarray}

When the scalar perturbation settles down, the
Einstein equations in
Eq.(\ref{equationsofmotion}) are reduced to
\begin{eqnarray}\label{Einsteineqn}
0&=&\left(1-\frac{2 \alpha
f(r)}{r^2}\right)f'(r)+\frac{2f(r)}{r}-\frac{4r}{l^2}+\frac{2}{3b^2}
\kappa ^2 r \left[\left(1-b^2
\phi'(r)^2\right)^{-1/2}-1\right],\nonumber\\
0&=&\left(1-\frac{2 \alpha  f(r)}{r^2}\right)
\chi '(r),
\end{eqnarray}
and  the equation governing the electromagnetic
field becomes
\begin{eqnarray}\label{bipotetial}
0&=&\left[\phi ''(r)+\left(\frac{\chi'(r)}{2
}+\frac{3}{r}\right)\phi '(r)\right](1-b^2 e^{\chi (r)}\phi '(r)^2 )
+\frac{b^2}{2} \phi '(r)^2 e^{\chi (r)}\left(\chi'(r)
\phi'(r)+2\phi''(r)\right).
\end{eqnarray}

We can do the rescales in Eq.(\ref{phifield}), (\ref{Einsteineqn})
and(\ref{bipotetial})
\begin{equation}\label{symmetry4}
e^{\chi} \rightarrow a^{2} e^{\chi},~\phi \rightarrow a^{-1}\phi,~t
\rightarrow a t,~\omega\rightarrow a^{-1} \omega,
\end{equation}
\begin{equation}\label{symmetry5}
l\rightarrow a l,~\alpha\rightarrow a^2 \alpha,~\omega\rightarrow
a^{-1} \omega,~r\rightarrow ar,~~q\rightarrow q /a,~m^2\rightarrow
m^2/a^{2},~b^2\rightarrow a^2 b^2 ,~k\rightarrow a^{-1} k
\end{equation}
\begin{equation}\label{symmetry6}
r\rightarrow ar, f\rightarrow a^2 f,~\phi\rightarrow
a\phi,~\omega\rightarrow a \omega,
\end{equation}
\begin{equation}\label{symmetry7}
q \rightarrow aq,~\phi \rightarrow a^{-1}\phi,~\kappa^2 \rightarrow
\kappa^2 a^{2},~b^2\rightarrow a^2 b^2.
\end{equation}
The first rescale guarantees $\chi(r)=0$, the second one sets AdS
radius as unity. Using the third rescale, we can set $r_h=1$. The
last rescale relates the parameter of the backreaction $\kappa^2$ to
the charge of the scalar field $q$. From (\ref{symmetry7}), we can
see that, as $\kappa^2$ is fixed the increase of $q$ corresponds to
the decrease of $\kappa^2$ with fixed $q$. Thus when $\kappa^2$ is
fixed, a larger $q$ leads to weaker backreaction.

Substituting $\chi(r)=0$, Eq.(\ref{bipotetial}) becomes
\begin{equation}\label{phieq2}
0=b^2 \phi'(r)^2 \phi''(r)+\left(1-b^2 \phi'(r)^2\right)
\left(\frac{3 \phi'(r)}{r}+\phi''(r)\right),
\end{equation}
which provides the analytical solution \cite{deycai,caipangwang}
\begin{equation}\label{phiso2}
\phi(r)=U+\frac{ Q
}{r^{2}}\,_{2}F_{1}\left[\frac{1}{3},\frac{1}{2},\frac{4}{3},-\frac{4b^2
Q^2}{r^6}\right],
\end{equation}
where $Q$ and $U$ are integral constants. $U$ can
be eliminated by the constraint at the horizon,
$\phi(r_h)=0$,
\begin{equation}
U=-Q
\,_{2}F_{1}\left[\frac{1}{3},\frac{1}{2},\frac{4}{3},-4b^2Q^2\right].\nonumber
\end{equation}
At infinity, Eq.(\ref{phiso2}) asymptotically behaves as
\begin{equation}
\phi(r)\sim U-\frac{Q}{ r^2}.\nonumber
\end{equation}

In the limit $b\rightarrow0$, Eq.(\ref{phieq2}) is reduced to
\begin{equation}\label{phieq3}
0=\frac{3 \phi'(r)}{r}+\phi''(r)
\end{equation}
which is the equation of motion of the electromagnetic field in the
Maxwell theory. The solution  reads
\begin{eqnarray}
\phi(r)=Q-\frac{Q}{ r^2}.\nonumber
\end{eqnarray}
Substituting  Eq.(\ref{phiso2}) into
Eq.(\ref{Einsteineqn}), the gravitational field
equation is rewritten as
\begin{equation}\label{ffun}
0=\left(1-\frac{2 \alpha f(r)}{r^2}\right)f'(r)+\frac{2
f(r)}{r}-\frac{4 r}{l^2}-\frac{2 \kappa^2 r}{3 b^2}+\frac{2
\kappa^2}{3b^2r^2}\sqrt{4 b^2 Q^2+r^6},
\end{equation}
which has an analytical solution \cite{wil,Miskovic}
\begin{eqnarray}\label{gbbi}
f(r)=\frac{r^2}{2\alpha}(1-\sqrt{g(r)}),
\end{eqnarray}
where the function $g(r)$ is
\begin{eqnarray}
g(r)=1-\frac{4\alpha}{l^2}+\frac{4\alpha
\widetilde{m}}{r^4}-\frac{\kappa^2
\alpha}{3b^2}+\frac{\kappa^2\alpha}{3b^2 r^3}\sqrt{r^6+4b^2
Q^2}-\frac{2\alpha \kappa^2 Q^2}{r^6} \,_2 F_1
[\frac{1}{3},\frac{1}{2},\frac{4}{3},-\frac{4 b^2 Q^2}{r^6}].
\end{eqnarray}
$\,_2F_1$ represents a hypergeometric function, $Q$ is an integral
constant which actually plays the role of electric charge of the
black hole and $\widetilde{m}$ is an integral constant relating to
the black hole ADM mass. In the limit of vanishing Gauss-Bonnet
factor, the background spacetime reduces to a Born-Infeld AdS black
hole \cite{deycai}. If we take the limit $b\rightarrow0$, the
background spacetime reduces to a Gauss-Bonnet Reissner Nordstrom
AdS black hole. There is an upper bound of the electric charge $Q_c$
contained inside the black hole, which indicates the phase
transition point of the black hole. Beyond $Q_c$, the black hole
will become unstable. This $Q_c$ has to be calculated numerically
for the Gauss-Bonnet Born-Infeld AdS black hole.

In order to investigate the scalar perturbation
Eq.(\ref{phifield}), near the horizon $r\sim1$,
we should impose the incoming wave boundary
condition
\begin{equation}\label{incoming}
\psi_{\omega,k}(r)\sim(r-1)^{- i\frac{\omega}{4\pi T}}.
\end{equation}
Introducing a new variable $\varphi$ as
$\psi_{\omega,k}(r)=\mathcal{\Re}(r)\varphi_{\omega,k}(r)$ and
choosing
$\mathcal{\Re}(r)=exp[-i\int^{r}_{1}\frac{\omega+q\phi(r)}{f(r)}]$,
which asymptotically approaches Eq.(\ref{incoming}) at the horizon,
one can express the boundary condition at the horizon as
$\left.\varphi_{\omega,k}\right|_{r=1}=const.$, and Eq.
(\ref{phifield}) then becomes
\begin{eqnarray} \label{main equation}
0&=&\varphi_{\omega,k} ''(r)+B_1(r) \varphi_{\omega,k} '(r)+B_2(r)
\varphi_{\omega,k}(r)~,
\end{eqnarray}
with
\begin{eqnarray}
B_1(r)&=&\frac{f'(r)}{f(r)}-\frac{2 i[q\phi (r)+\omega]}{
f(r)}+\frac{3}{r}~,\nonumber\\
B_2(r)&=&-\frac{m^2}{f(r)}-\frac{k^2}{r^2 f(r)}-\frac{i [3(q\phi
(r)+\omega)+r \phi'(r)]}{r f(r)}~.
\end{eqnarray}

Near the AdS boundary $r\sim \infty$, $\varphi_{\omega,k}$ behaves
as
\begin{equation}\label{asymptotic behavior}
\varphi_{\omega,k}(r)\sim
\frac{\varphi_{\omega,k}^{-}}{r^{\lambda_{-}}}+\frac{\varphi_{\omega,k}^{+}}{r^{\lambda_{+}}}.
\end{equation}
where $\lambda_{\pm}=2\pm\sqrt{4+m^2 l_e^{2}}$
are characteristic exponents of the perturbation
equation, and $l_e ^2$ is the effective AdS
radius. The boundary conditions at the horizon
are now given by
\begin{eqnarray}\label{boundary}
\varphi_{\omega,k}|_{r=1}&=&1~,\nonumber\\
\left.\frac{\varphi^{\prime}_{\omega,k}}{\varphi_{\omega,k}}\right|_{~r=1}
&=&-\left.\frac{B_2(r)}{B_1(r)}\right|_{r=1}~.
\end{eqnarray}
Eq. (\ref{main equation}) is a linear equation and
$\varphi_{\omega,k}(r)$ must be regular at the horizon. Since we do
not concentrate on the amplitude of $\varphi_{\omega,k}(r)$, we can
set $\left.\varphi_{\omega,k}\right|_{r=1}=1$. Eq. (\ref{main
equation}) has to be solved numerically under the boundary
conditions (\ref{boundary}) and $\varphi_{\omega,k}^{-}=0$ at
infinity.

We will examine the behavior of the charged
scalar field perturbation which can present us an
objective picture on how the black hole evolves
when the temperature drops and when the
backreaction becomes stronger. Without loss of
generality, hereafter we will set $m^2=-3/l^2$ in
our calculation.

\subsection{Dynamical behavior of the scalar perturbation in the bulk}

Hereafter we will report dynamical perturbation
and critical phenomenon. We will pay more
attention on how the Born-Infeld factor, the
Gauss-Bonnet factor and the backreaction
influence the perturbation in the bulk background
spacetime when the temperature of the black hole
drops. Further we are going to study the critical
phenomenon once the black hole approaches
marginally stable.

For the scalar perturbation, we will concentrate on the lowest
quasinormal mode as studied in
\cite{wang,liupanwang,zhuwang,wangMolina}. We can obtain the
quasinormal frequencies by solving Eq.(\ref{main equation}) based on
the boundary conditions (\ref{boundary}) at the horizon and
$\varphi_{\omega,k=0}^{-}=0$ at the AdS boundary. For a stable
spacetime, the outside perturbation will exhibit a decay mode. When
the perturbation grows up, the background spacetime becomes
unstable.

As the background spacetime approaches marginally stable and finally
changes to be unstable, the black hole will experience a phase
transition. The correlation length is the scale parameter that
exists in the system near the phase transition point. It increases
while the temperature approaches its critical value and becomes
infinite at the moment of the phase transition. The correlation
length is defined as $\xi^{2}:=-k^{-2}$ \cite{maeda} and can be
calculated by solving Eq. (\ref{main equation}) with $\omega=0$
under boundary conditions Eq. (\ref{boundary}) at the horizon and
$\varphi_{\omega=0,k}^{-}=0$ at the AdS boundary.

\begin{figure}[h]
\includegraphics[width=200pt]{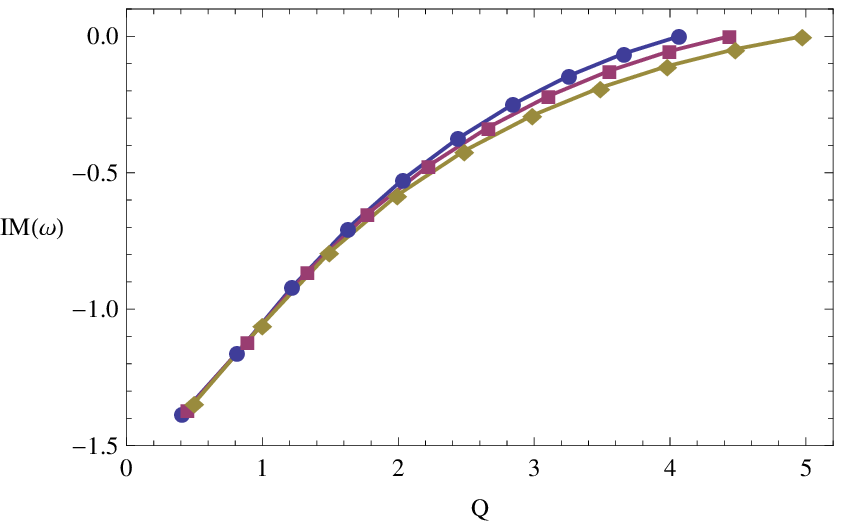}
\includegraphics[width=188pt]{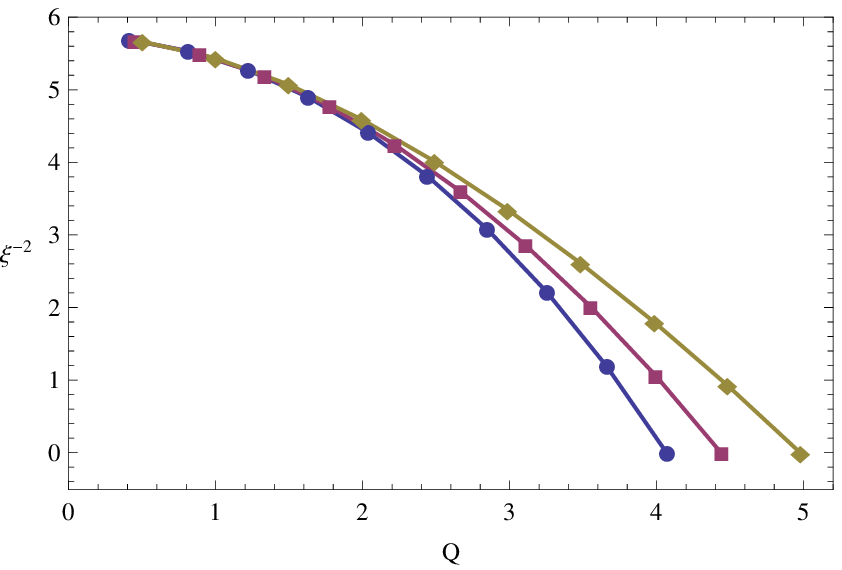}
\caption{\label{qnm2} (Color online) Left) The
imaginary part of the lowest quasinormal
frequency for different values of $b$ when
$\kappa^2$ and $\alpha$ are fixed to be 0.05 and
0 respectively.  Lines from top to bottom
correspond to $b=0.05, 0.1$ and $0.14$,
respectively. Right) The correlation length $\xi$
as a function of $Q$ for different values of $b$.
Lines from bottom to top correspond to
$b=0.05,0.1$ and $0.14$, respectively. }
\end{figure}

Let us first see the influence on the quasinormal
modes and correlation length given by the
Born-Infeld factor. To illustrate the influence,
we set a vanishing Gauss-Bonnet factor and fix
the backreaction parameter $\kappa^2=0.05$. In
the left panel of Fig.\ref{qnm2} we show the
dependence of the imaginary part of quasinormal
frequency on the Born-Infeld factor $b$. With the
increase of the electric charge of the black hole
$Q$, the imaginary part of the quasinormal
frequency approaches  zero monotonously. When
$Q<Q_c$, the imaginary part of the frequency
keeps negative, which shows that the perturbation
has a decay mode ensuring the stable spacetime.
For the fixed $Q$, we observe that a bigger value
of the Born-Infeld factor $b$ leads to bigger
absolute value of the imaginary part of the
quasinormal frequency. Considering that the decay
time scale of the perturbation $\tau\sim
1/|IM(\omega)|$, bigger $b$ will make the
perturbation decay faster. This means that the
stronger nonlinearity in the electromagnetic
field can make the perturbation die out quicker
and protect the background spacetime to be more
stable.

The influence of Born-Infeld factor $b$ on the
correlation length is shown in the right panel of
Fig. \ref{qnm2}. As $Q$ tends to its critical
value $Q_c$, the correlation length approaches
infinity. A larger $b$ slows down the process to
make $\xi$ infinity, which implies that the black
hole background is more stable when $b$ is
bigger. This is consistent with the observation
in the quasinormal modes behavior.

\begin{figure}[h]
\includegraphics[width=200pt]{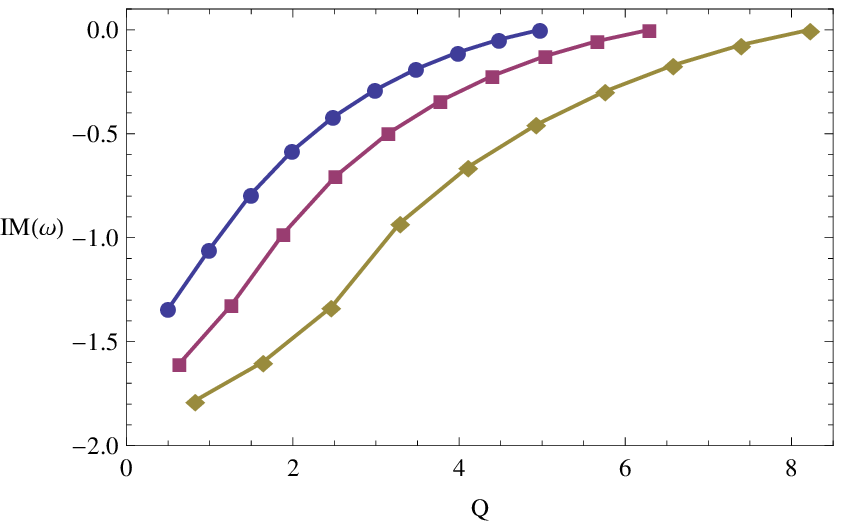}
\includegraphics[width=188pt]{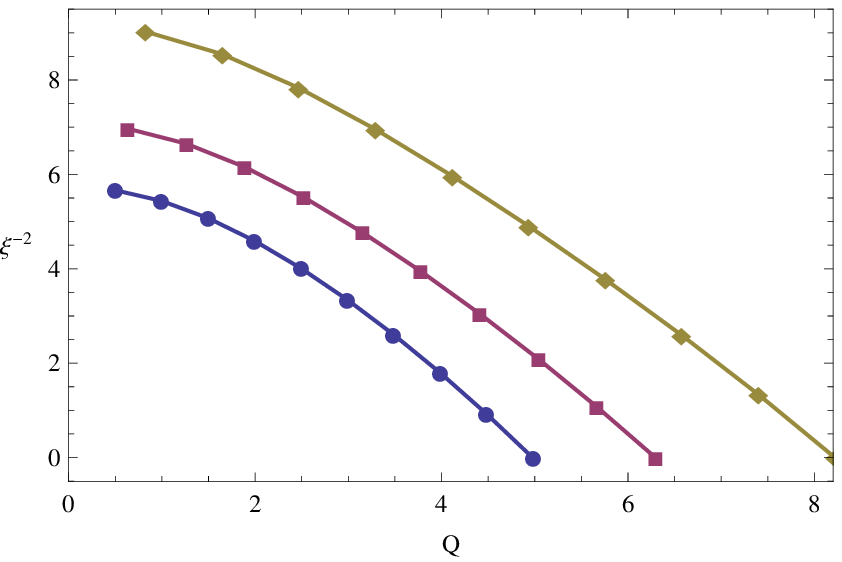}
\caption{\label{qnm3} (Color online) Left) The
imaginary part of the lowest quasinormal
frequency for different values of $\alpha$ while
$b^2$ and $\kappa^2$ are fixed to be 0.02 and
$0.05$ respectively. Lines from  top to bottom
correspond to $\alpha=0, 0.1$ and $0.2$,
respectively.  Right) The correlation length
$\xi$ as a function of $Q$ for different values
of $\alpha$. Lines from bottom to top correspond
to $\alpha=0,0.1$ and $0.2$, respectively.}
\end{figure}

Now we turn to exhibit the influence of the Gauss-Bonnet factor on
the perturbation in the background of Gauss-Bonnet Born-Infeld AdS
black hole. We fix the backreaction factor $\kappa^2$ and the
Born-Infeld factor $b$ to be 0.05 and 0.14, respectively, in our
numerical calculation. As shown in Fig.\ref{qnm3}, the imaginary
part of quasinormal frequency keeps approaching zero with the
increase of the black hole charge $Q$. For the fixed black hole
charge, we observe that when the Gauss-Bonnet factor $\alpha$ is
bigger, the absolute value of the imaginary part of the quasinormal
frequency becomes bigger. This tells us that the bigger Gauss-Bonnet
factor makes the perturbation decay quicker, which corresponds to
say that the bigger $\alpha$ can keep to background spacetime to be
more stable.

The dependence of correlation length of scalar
perturbation on the Gauss-Bonnet factor is shown
in the right panel of Fig.\ref{qnm3}. As the
electric charge of black holes approach their
critical values, the correlation length keeps
increasing to infinity. A larger $\alpha$ can
slow down the process to make $\xi$ infinity,
which means that more curvature correction can
delay the system to approach the phase transition
point. This is consistent with the quasinormal
behavior, indicating that the background is more
stable when the $\alpha$ is bigger.

From the above results, we find that the higher order gravity
correction and the higher correction to the gauge matter field
basically play the same role in the perturbation in the background
Gauss-Bonnet Born-Infeld AdS black hole. Both of them can protect
the spacetime to be away from instability.

\begin{figure}[h]
\includegraphics[width=200pt]{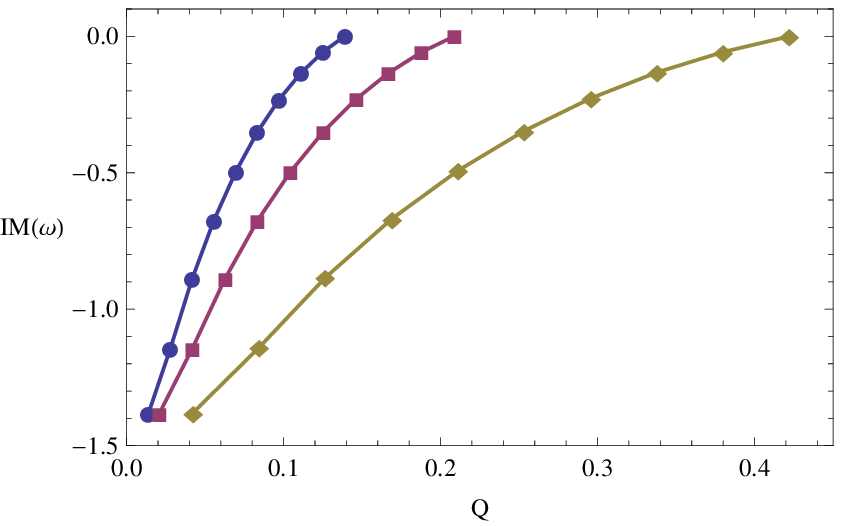}
\includegraphics[width=188pt]{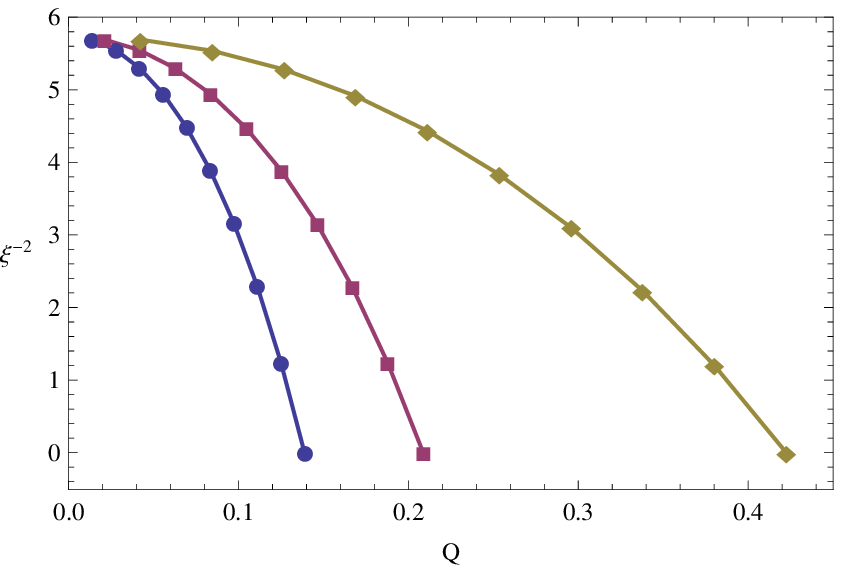}
\caption{\label{qnm1} (Color online) Left) The imaginary parts of
the lowest quasinormal frequency for different values of $q$ while
$\kappa^2$,$b$ and $\alpha$ are fixed to be 1, 0.1 and 0
respectively. Lines from bottom to top correspond to $q=10, 20$ and
$30$, respectively. Right) The correlation length $\xi$ as a
function of $Q$ for different $q$. Line from top to bottom
correspond to $q=10,20$ and$30$, respectively.}
\end{figure}

In order to investigate the influence of the backreaction, one has
to fix $\kappa^2$, say $\kappa^2=1$, and study how the parameter $q$
affects the behavior of the scalar perturbation. Using the rescale
(\ref{symmetry7}) to set $q=1$ will change $Q$ in different scale
for various strength of the backreaction. It is worth noting that
the larger the $q$ is, the weaker the backreaction is. Fig.
\ref{qnm1}  shows the dependence of perturbation frequency and
correlation length with the change of the coupling $q$ while
backreaction $\kappa^2$, Born-Infeld factor $b$ and Gauss-Bonnet
factor $\alpha$ are fixed to be 1, 0.1 and 0 respectively. The
frequencies of the scalar perturbation approach zero as the electric
charge approaches to its critical value. The stronger coupling $q$
makes $IM(\omega)$ approach zero quicker. This indicates that the
black hole background is easier to be destroyed when the
backreaction is weaker and it is more stable when the backreaction
is stronger. The behavior of the correlation length further supports
this result. The weaker coupling $q$ (stronger backreaction) makes
$\xi$ increase slower, which shows that the background is more
stable as the backreaction increases.

\subsection{The critical temperature in the bulk spacetime and its dependence on Gauss-Bonnet and Born-Infeld parameters}

The charge density $Q$ determines the phase
structure. When we fix parameters
$\kappa^2,b,\alpha$ and $q$, the critical value
of the black hole charge $Q_{c}$ indicates the
starting point of the phase transition. This
$Q_c$ can be obtained by solving Eq.(\ref{main
equation}) with $\omega=k=0$ under the boundary
condition at the horizon and
$\varphi_{\omega=0,k=0}^{-}=0$ at infinity using
the shooting method. Inserting the value of $Q_c$
into the Hawking temperature,  we can obtain the
critical temperature $T_c$ in the bulk black hole
spacetime.

In the following two tables, \ref{table1} and \ref{table2}, we list
the values of $T_c$ in a Gauss-Bonnet Born-Infeld AdS black hole
described by Eq.(\ref{temperature}) with different values of
parameters. From table \ref{table1} we can see that $T_c$ decreases
with the increase of the Born-Infeld factor regardless of the
choices of the Gauss-Bonnet factor. In table \ref{table2}, the
Born-Infeld factor is fixed as $0.14$, we find that $T_c$ decreases
with the increase of $\alpha$. This shows that both the higher order
gravity correction and the higher correction to the gauge matter
field  can make the critical temperature smaller, which ensures the
black hole background to be more stable.

\begin{center}
\begin{table}[ht]\label{table1}
\caption{\label{table1} The dependence of the critical temperature
$T_c/Q_c^\frac{1}{3}$ on $b$ for different values $\alpha$. We set
$q=1$, $\kappa^2=0.05$ and $m^2 l^2=-3$ in our numerical
computation. }
\begin{tabular}{|c|c|c|c|}
\hline
\backslashbox{$\alpha$}{$b$} & 1/20  & 1/10  &  $\sqrt{2}$/10  \\
[0.5ex] \hline $0$ & 0.147&~0.139~&0.130\\ \hline
 1/10 &~&~0.110~&~0.0976~\\ \hline
\end{tabular}
\end{table}
\end{center}
\begin{center}\label{table2}
\begin{table}[ht]
\caption{\label{table2} The dependence of the critical temperature
$T_c/Q_c^\frac{1}{3}$ on the Gauss-Bonnet factor $\alpha$. We set
$q=1$, $\kappa^2=0.05$, $b=0.14$ and $m^2=-3/l^2$ in the numerical
computation.}
\begin{tabular}{|c|c|c|c|c|c|}
\hline
 $\alpha$&-0.1 & 0 &  0.1 &  0.15  &  0.2  \\
[0.5ex] \hline
$T_c/Q_c^\frac{1}{3}$ &~0.156~&~0.130~&~0.0976~&~0.0782~&~0.0571~\\
\hline
\end{tabular}
\end{table}
\end{center}

\section{Solutions at Low temperature phase}

When the black hole temperature drops to $T_c$, we observed that the
imaginary part of the quasinormal frequency will change from
negative to zero, which means that a stable black hole background
will become marginally stable. When $T\sim T_c$, the original
Gauss-Bonnet Born-Infeld AdS black hole spacetime will be destroyed
and a new configuration with scalar hair will be formed.  In order
to study the condensation of the scalar hair,  we will use the
shooting method to solve the equations of motions
(\ref{equationsofmotion}) numerically with appropriate boundary
conditions.

At the black hole horizon $r_h$, which is the
root of $f(r_h)=0$, the solutions of the gauge
and the scalar fields have to be regular
\begin{eqnarray}\label{horizonboundry}
\phi(r_h)=0~,  ~~\psi'(r_h)=\frac{m^2}{f'(r_h)}\psi(r_h),
\end{eqnarray}
and the coefficients in the metric ansatz obey
\begin{eqnarray}\label{horizonmetric}
f'(r_h)&=&\frac{4r_h}{l^2}-\frac{2}{3} \kappa ^2 r_h \left[m^2 \psi
(r_h)^2\frac{e^{\chi (r)} q^2 \phi(r)^2
\psi(r)^2}{f(r)}+\frac{1}{b^2}\left[\left(1-b^2
\phi'(r_h)^2\right)^{-1/2}-1\right]\right],\\
\chi '(r_h)&=&-\frac{4}{3} \kappa ^2 r_h \left[\frac{q^2 \phi'
(r_h)^2 \psi (r_h)^2 e^{\chi (r_h)}}{f'(r_h)^2}+\psi '(r)^2\right].
\end{eqnarray}
At the spatial infinity, the asymptotic behaviors of the solutions are
\begin{eqnarray}\label{infinity}
&&\chi\rightarrow 0~,~~~f(r)\sim \frac{r^2}{l^2}~,\nonumber\\
&&\phi(r)\sim \mu-\frac{\rho}{r^2}~,~~~\psi(r)\sim
\frac{\psi_-}{r^{\lambda-}}+\frac{\psi_+}{r^{\lambda+}},
\end{eqnarray}
where $\lambda_\pm=2\pm\sqrt{4+m^2 l_e^2}$, $\mu$ is the chemical
potential. We choose $\psi_-=0$, thus one can interpret
 $\langle\mathcal{O}_{\lambda+}\rangle=\psi_+$, where
 $\mathcal{O}_{\lambda+}$ is the operator with the conformal
 dimension $\lambda+$ dual to the scalar field.

There are several scaling symmetries to be adopted in the above
equations as those in \cite{Gregory,3h2}
\begin{equation}\label{symmetry0}
e^{\chi} \rightarrow a^{2} e^{\chi} ,~\phi \rightarrow \phi/ a,~t
\rightarrow a t,
\end{equation}
\begin{equation}\label{symmetry1}
l\rightarrow a l,~\alpha\rightarrow a^2
\alpha,~r\rightarrow ar,~~q\rightarrow q
/a,~m^2\rightarrow m^2/a^{2},~b^2\rightarrow a^2
b^2 ,
\end{equation}
\begin{equation}\label{symmetry2}
r\rightarrow ar, f\rightarrow a^2 f,~\phi\rightarrow a\phi,
\end{equation}
\begin{equation}\label{symmetry3}
q \rightarrow aq,~\phi \rightarrow \phi/a,~\psi
\rightarrow \psi/ a,~\kappa^2 \rightarrow
\kappa^2 a^{2},~b^2\rightarrow a^2 b^2.
\end{equation}
We can use symmetry (\ref{symmetry0}) to set the
asymptotic value of $\chi$ to zero. Employing
symmetry (\ref{symmetry1}), we have  $l=1$.
Further using symmetries
$(\ref{symmetry2}),(\ref{symmetry3})$, we get
$r_h=1$ and $q=1$.

The behavior of the holographic superconductor
can be found by numerically integrating
Eq.(\ref{equationsofmotion}) with the appropriate
boundary conditions
Eq.(\ref{horizonboundry})-Eq.(\ref{infinity}) at
the horizon and the AdS boundary.

In the following we will discuss the scalar
condensation in the holographic superconductor
and disclose its dependence on the parameters of
the backreaction, Gauss-Bonnet factor and
Born-Infeld factor. The numerical computation
allows a full exploration of the condensation of
the scalar field
$\langle\mathcal{O}_{\lambda+}\rangle$.

\begin{figure}[h]\label{alloperator}
\includegraphics[width=220pt]{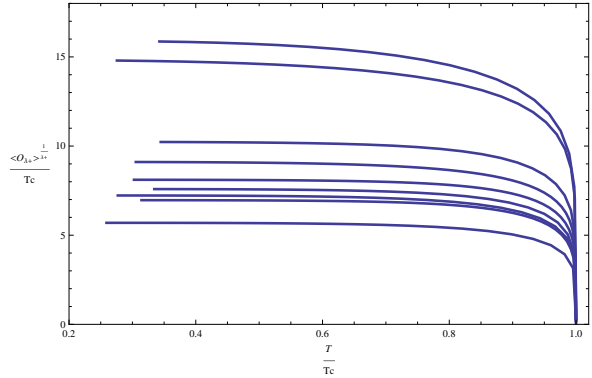}
\caption{\label{alloperator} Dependence of the
scalar  condensate on various parameters.}
\end{figure}

The dependence of the scalar condensation on
various parameters is shown in
fig.\ref{alloperator}. For the convenience of the
following discussion, we number lines from bottom
to top in fig.\ref{alloperator} in the order of 1
to 9. The physics parameters of each line are
listed in Table \ref{table5}. We observe that
when the condensation gap becomes bigger,  the
critical temperature to start the condensation
becomes smaller, which means that it is more
difficult for the scalar hair to condense. The
critical temperature for the condensation of the
scalar operator calculated in CFT coincides with
that reported in  tables \ref{table1} and
\ref{table2} from the analysis of the spacetime
stability in the bulk. This gives further
evidence of the holography.

\begin{center}\label{table5}
\begin{table}[h]
\caption{\label{table5} This table provide the
characteristic parameter of each line in
Fig.\ref{alloperator}. The lines in
Fig.\ref{alloperator} have been labeled from 1 to
9 from bottom to top.}
\begin{tabular}{|c|c|c|c|c|c|c|c|c|c|}
\hline
$line$ & 1 &2&3&4&5&6&7&8&9\\
\hline
$T_{c}/\rho^\frac{1}{3}$  &0.189&0.156&0.147&0.139&0.130&0.110&0.0976&0.0582&0.0571\\
\hline
$\alpha$ &0 &-0.1 &  0 &  0  &  0&0.1&0.1&0.1&0.2 \\
\hline
$\kappa^2$ &0 &0.05 & 0.05 & 0.05 & 0.05&0.05&0.05&0.1&0.05\\
\hline
$b$ &0.1 &0.14 & 0.05 & 0.1 & 0.14&0.1&0.14&0.14&0.14\\
\hline
\end{tabular}
\end{table}
\end{center}

We now discuss the dependence of the condensation on various
parameters. In lines 3, 4,and 5, we take the same values of the
Gauss-Bonnet factor and the strength of the backreaction, namely
$\alpha=0, \kappa^2=0.05$, we see that the condensation gap
increases as the Born-Infeld factor increases.  This means that the
scalar condensation becomes harder to be formed when the
nonlinearity in the electromagnetic filed increases. This agrees
with the observation that the stability is kept well with the highly
nonlinear electromagnetic field observed in the bulk dynamics study.
This property also holds when we compare lines 6 and 7.

If we look at lines 2, 5, 7 and 9, where the strength of the
backreaction and the Born-Infeld factor are fixed to be
$\kappa^2=0.05, b=0.14$, we observe that the condensation gap
increases with the increase of the Gauss-Bonnet factor. Comparing
with the result in \cite{Pan-Wang}, we find that the higher order
gravity correction influence on the condensation will not be changed
when the scalar field is coupled to the nonlinear electromagnetic
field. The Gauss-Bonnet factor will make the condensation harder for
the scalar field no matter whether the scalar field is coupled to a
linear or nonlinear gauge matter field.

The effect on the condensation brought by the backreaction is
similar to the corrections in gravity and gauge matter field. The
condensation gap increases when the backreaction becomes stronger,
which indicates that the condensation becomes harder to form when
there is stronger backreaction from the matter fields to the
gravity. This is also consistent with the observation in the
background stability analysis.

The influences given by the Born-Infeld,
Gauss-Bonnet and backreaction factors can be
further understood by examining the effective
mass of the scalar field, which is expressed as
\begin{equation}
m^2_{eff}=m^2+q^2 A^2_t
g^{tt}(r)=-\frac{3}{l^2}-\frac{\phi^2(r)}{f(r)}e^{\chi(r)}.
\end{equation}
$g^{tt}(r)$ is negative, so there is a chance that the effective
mass becomes sufficiently negative near the horizon to destabilize
the scalar field. From Fig.\ref{meff} we observe that with the
growth of the Born-Infeld factor and the strength of the
backreaction, the effective mass develops a more shallow well out of
the horizon. Similar phenomenon also happens when we increase the
Gauss-Bonnet factor. The negative effective mass is the crucial
effect to cause the formation of the scalar hair and the less
negative effective mass will make it harder for the scalar hair to
form \cite{gubser}. This picture explains the reason why the
stronger backreaction, the bigger corrections in gravity and gauge
matter field can make the scalar condensation develop harder.
\begin{figure}[h]\label{meff}
\includegraphics[width=195pt]{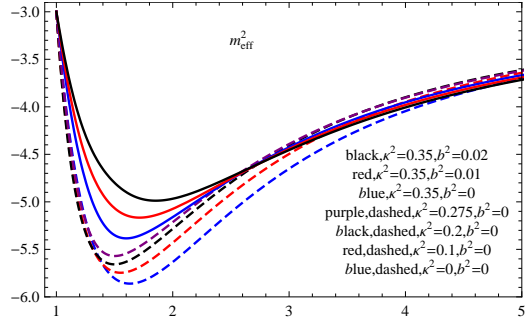}
\caption{\label{meff} The dependence of $m_{eff}^2$ on different
parameters with $\psi_+= 4$ in all the plots.}
\end{figure}

\section{Conductivity}

In order to investigate the influences of the
backreaction, the Born-Infeld factor and the
Gauss-Bonnet factor on the conductivity, we
consider the time-dependent perturbations with
zero momentum $A_i(t,r,x^i)=A(r)e^{-i \omega
t}e_i$~ and $h_{ti}(t,r,x^i)=h_{ti}(r)e^{-i
\omega t}$, where $A_i$ is the perturbation of
the vector potential which has only spatial
components and $h_{ti}(t,r,x^i)$ is the
perturbation of the metric tensor. After some
algebra, we obtain the linear equation of motion.

The perturbation equation for $A(r)$ reads
\begin{eqnarray}\label{ax0}
&&-r \phi '(r) e^{\frac{\chi (r)}{2}}\left[ h_{ti}
'(r)-\frac{2}{r}h_{ti}(r)\right]\nonumber\\
&&=\left[e^{-\frac{\chi (r)}{2}}r f(r)
A'(r)\right]'+\frac{A(r)e^{\frac{\chi (r)}{2}}\omega^2 r}{f(r)}-2
q^2 r \psi(r)^2 A(r)e^{-\frac{\chi (r)}{2}}\left[1-b^2 e^{\chi (r)}
\phi
   '(r)^2\right]^\frac{1}{2}\nonumber\\
   &&\frac{1}{2} b^2 r e^{\frac{\chi (r)}{2}} A'(r) f(r) \left[\phi
'(r)^2 \chi '(r)+2 \phi '(r) \phi ''(r)\right]\left[1-b^2 e^{\chi
(r)} \phi
   '(r)^2\right]^{-\frac{1}{2}}.
\end{eqnarray}
The perturbation equation for $h_{ti}(r)$ has the
form
\begin{eqnarray}
\left[1-b^2 \phi '(r)^2 e^{\chi
(r)}\right]^\frac{1}{2} \left[h_{ti}
'(r)-\frac{2}{r}h_{ti}(r)\right]+\frac{2 \kappa^2
r^2 A(r)\phi '(r)}{r^2-2 \alpha f(r) }=0.
\end{eqnarray}
By substituting the above equation into Eq.(\ref{ax0}), we have
\begin{eqnarray}\label{ax}
0&=&\frac{1}{2} b^2 e^{\chi (r)} A'(r) \left[\phi '(r)^2 \chi '(r)+2
\phi '(r) \phi ''(r)\right]+\left[1-b^2 e^{\chi (r)} \phi
   '(r)^2\right] \left[A''(r)+A'(r) \left[\frac{f'(r)}{f(r)}-\frac{\chi '(r)}{2}+\frac{1}{r}\right]\right.\nonumber\\&&\left.+\frac{\omega ^2 A(r)
   e^{\chi (r)}}{f(r)^2}\right]-A(r) \left[\frac{2 \kappa ^2 r^2 e^{\chi (r)} \phi '(r)^2 \left[1-b^2 e^{\chi (r)} \phi
   '(r)^2 \right]^\frac{1}{2}}{f(r) \left[r^2-2 \alpha  f(r)\right]}+\frac{2 q^2 \psi (r)^2 \left[1-b^2 e^{\chi (r)} \phi
   '(r)^2\right]^{\frac{3}{2}}}{f(r)}\right]
\end{eqnarray}
This equation is solved under the boundary
condition of incoming wave at horizon
\begin{eqnarray}
A(r)\sim f(r)^{-i\frac{\omega}{4\pi T_+}}~,
\end{eqnarray}
where $T_+$ is the temperature of the $(hairly)$
black hole. At infinity the asymptotic behavior
of $A(r)$ is
\begin{eqnarray}
A(r)=a_0+\frac{a_2}{r^2}+\frac{a_0 l_e^4 \omega^2}{2
r^2}log\frac{r}{l}~,
\end{eqnarray}
where $a_0$ and $a_2$ are integration constants.
Then the conductivity is found to
be\cite{Barclay-Gregory}
\begin{eqnarray}
\sigma=\frac{2 a_2}{i \omega L_2^4 a_0}+\frac{i
\omega}{2}-i \omega log(\frac{l_e}{l}).
\end{eqnarray}

In the following figures we show the numerical
results of conductivity and its dependence on
different parameters. The blue line represents
the real part and the purple line shows the
imaginary part of $\sigma$ at temperature
$T/T_c\simeq0.32$.

We observe that the influence of the Gauss-Bonnet factor still keeps
the same as that disclosed in \cite{Pan-Wang} when the gauge matter
field is not the usual Maxwell field. With the increase of the
Gauss-Bonnet coupling constant, the gap frequency becomes larger.
The nonlinearity in the gauge field does not change this property.
With the increase of the Born-Infeld factor, we observe that the
conductivity gap frequency becomes larger as well. Thus the higher
order gravity correction and the higher correction to the gauge
matter field basically play the same role in the conductivity.
Similar influence has also been found for the blackreaction.

\begin{figure}[ht]\label{figure6}
\includegraphics[width=190pt]{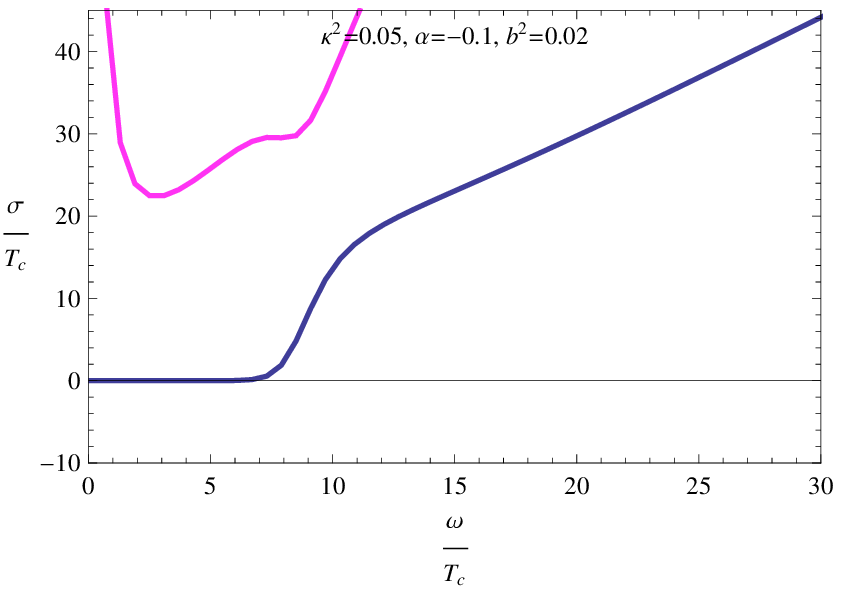}
\includegraphics[width=190pt]{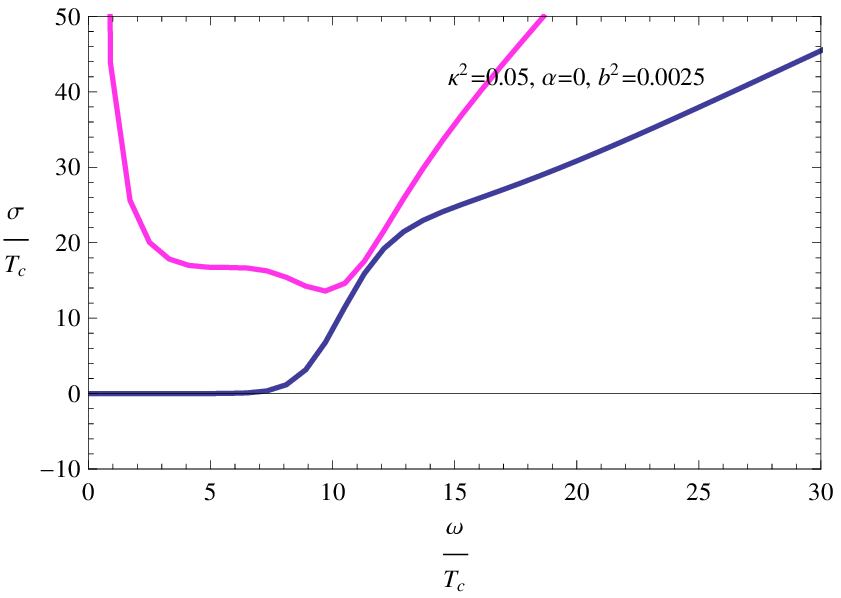}
\includegraphics[width=190pt]{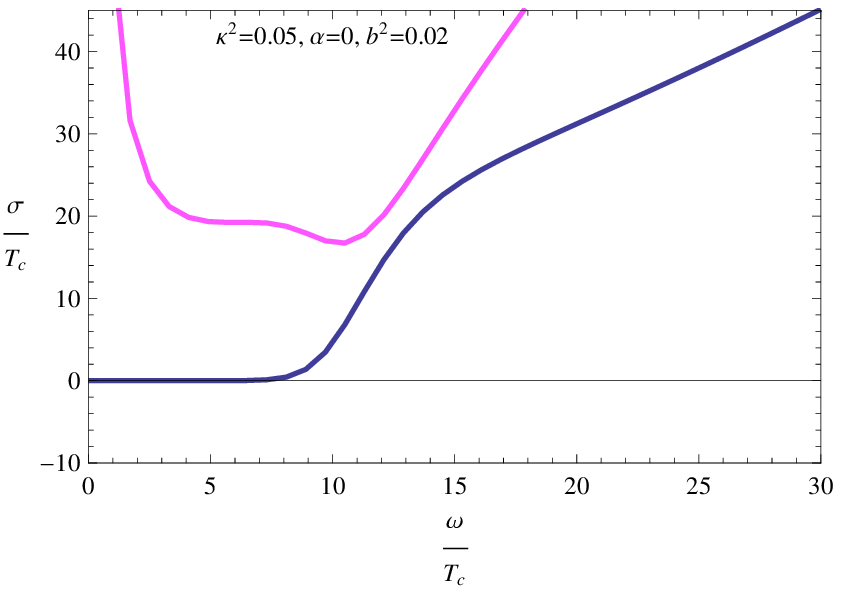}
\includegraphics[width=190pt]{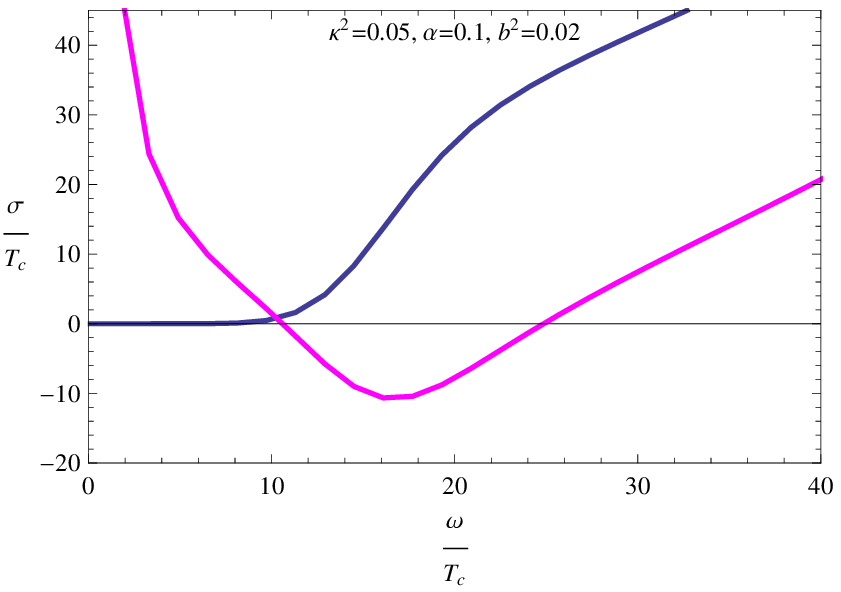}
\caption{\label{figure6} Influence on the
conductivity caused by the Gauss-Bonnet factor
and the Born-Infeld factor. }
\end{figure}

\section{Conclusions}

In this work we have studied a general picture of holographic
superconductor.  We have considered high curvature correction in the
gravity side and nonlinear correction in the gauge matter field.
Instead of the probe limit, we have taken the backreaction of
classical fields onto the spacetime into account.  With the
backreaction, we have obtained the Gauss-Bonnet Born-Infeld AdS
black hole. We have analyzed the dynamics of this gravitational
background in the bulk and examined the condensation of the scalar
hair in the boundary CFT.

We have obtained the consistent result that the correction in the
gravity can make the bulk background more stable and the scalar hair
condensation on the boundary more difficult to develop. This
property always holds no matter whether the scalar field is coupled
to a Maxwell field or a nonlinear electromagnetic field. Furthermore
we found that in addition to the neutral AdS black hole background,
in the more general configuration such as a nonlinearly charged
Gauss-Bonnet AdS black hole, the nonlinear correction in the gauge
matter field persists the similar effect in the background stability
and condensation of scalar field to that of the correction in the
gravity side. The nonlinearity of the electromagnetic field can
protect the stability of the background spacetime and make the
scalar hair condensation hard to form on the boundary. The
consistent effect brought by the corrections in gravity and gauge
matter field is different from that observed in calculating the the
ratio of shear viscosity to entropy density in the AdS background
\cite{caisun}. The reason behind this difference needs to be
clarified. The strength of the backreaction plays the similar role
to the corrections in gravity and gauge matter field on the
spacetime stability and scalar hair formation. The consistent
influences of the strength of the backreaction and corrections in
gravity and gauge matter field have also been disclosed in the
effective mass of the scalar field and the conductivity.

\begin{acknowledgments}

This work was partially supported by the National Natural Science
Foundation of China.

\end{acknowledgments}

\end{CJK}
\end{document}